\documentclass[12pt]{article}
\usepackage{latexsym}
\newcommand{\beq}{\begin{equation}} 
\newcommand{\eeq}{\end{equation}}
\newcommand{\beqs}{\begin{eqnarray}} 
\newcommand{\eeqs}{\end{eqnarray}}
\newcommand{\tr}{\mathrm{tr}}

\begin{document}
\begin{titlepage}
\vskip 2.5cm
\begin{center}
{\LARGE \bf Chiral $SU(N)$ Gauge Theories}\\
\smallskip
{\LARGE \bf and the Konishi Anomaly}\\
\vspace{2.71cm}
{\large
Riccardo Argurio,
Gabriele Ferretti and 
Rainer Heise}
\vskip 0.7cm
{\large \it Institute for Theoretical Physics - G\"oteborg University and \\
\smallskip
Chalmers University of Technology, 412 96 G\"oteborg, Sweden}
\vskip 0.3cm
\end{center}
\vspace{3.14cm}
\begin{abstract}
We study chiral $SU(N)$ supersymmetric gauge theories
with matter in the antifundamental and antisymmetric representations.
{}For $SU(5)$ with two families, we show how to reproduce the 
non-perturba\-ti\-ve\-ly generated superpotential, and we discuss dynamical
supersymmetry breaking purely in terms of the Konishi anomaly.
We apply the same technique in general to $SU(N)$ with one family. 
We also briefly comment on the chiral ring for these theories. 
\end{abstract}

\end{titlepage}

\section{Introduction}
In the context of ${\cal N}=1$ supersymmetric gauge theories, a very 
interesting possibility is to break
supersymmetry by non-perturbative dynamics. However, dynamical
supersymmetry breaking is only possible in a restricted class of theories,
the so-called chiral theories, i.e.
those with a matter content which does not allow to write a mass
term for all matter fields. 
Chiral theories with gauge group $SU(N)$ are the most well-known
examples of theories where dynamical supersymmetry breaking can occur.

That dynamical supersymmetry breaking does occur in a number of chiral
theories has been convincingly argued some time ago \cite{ads,mv}.
The two lines of arguments put forward are the following. One can 
determine the  
non-perturbatively generated superpotential by symmetry arguments, and show
that 
together with the tree-level superpotential it does not lead to an
extremum \cite{ads}. 
Alternatively, one can use the Konishi anomaly relations \cite{konishi} 
to prove
that in a supersymmetric vacuum the gluino condensate $S=-{1\over 32 \pi^2}
\tr W^\alpha W_\alpha$ must vanish, and then show with a one-instanton
calculation that $S$ is actually non-zero in the vacuum \cite{mv}, thus
contradicting the assumption of a supersymmetric vacuum.

The object of the present paper is to discuss
dynamical supersymmetry breaking by using the Konishi anomaly to determine
the non-perturbative piece of the effective superpotential. If such a term
is non-zero, one can then see whether the addition of a tree level
superpotential leads to a supersymmetric vacuum or not. If, on the other
hand, there is an insufficient number of gauge invariants,
the Konishi relations typically lead to a vanishing gluino 
condensate $S$. One then has to resort to the arguments of \cite{ads, mv}
for the case where no non-perturbative superpotential is expected, which
point towards a strongly coupled non-supersymmetric vacuum.

Although the introduction and use of the generalized Konishi anomaly 
\cite{cdsw} has recently led to some new insight in the dynamics of 
${\cal N}=1$ gauge theories, we only make use of the original Konishi
anomaly. In one instance we can use a 
slight generalization, which leads however only to a classical relation
(as in \cite{corley} for SQCD with baryonic deformations).
We use a simplified version of the method outlined in
\cite{binor}, where it was applied to one particular chiral theory 
(which does not have dynamical supersymmetry breaking). For a related
discussion see \cite{naka}. 

The generalization of the Konishi anomaly seems non-trivial for these
theories which do not have a matter field in
the adjoint like in \cite{cdsw} 
(or at least in a real representation with two indices, see \cite{kraus}).
Similarly, for the same reason (and because of the impossibility to write
mass terms for the matter fields) it seems difficult to have a matrix
model formulation of the chiral theories in the spirit of 
\cite{DV,proof}.

In the following, we will consider first of all the $SU(5)$ gauge theory
with 2 families of chiral matter in the $\bf 10$ and $\bf \bar 5$
representations, which is the most celebrated 
chiral theory where dynamical supersymmetry breaking occurs at arbitrarily
small coupling. We then proceed to consider more generally the 
$SU(N)$ theories with one antisymmetric and $N-4$ antifundamental (i.e. one 
family), which can have a supersymmetric vacuum for $N$ even while
they break supersymmetry dynamically (but at strong coupling) for $N$ odd.
Finally we show that for
$N$ odd and one family the chiral ring 
is generated only by the glueball superfield and
by the basic invariants.

\section{$SU(5)$ chiral gauge theory with two families}
The matter content
of this $SU(5)$ gauge theory is given by 2 flavors
of anti-fundamentals $\tilde F^{\tilde \imath}_a$ and 2 flavors of
antisymmetric tensors $T^{ab}_{i}$ \cite{ads,mv}. 
The matter content is of course such 
that the gauge anomaly cancels. The global symmetries of the theory
are at the classical level $SU(2)_{\tilde F}\times SU(2)_T \times
U(1)_{\tilde F}\times U(1)_T \times U(1)_R$. At the quantum level, the
$U(1)$s are anomalous, but 2 anomaly-free combinations can always be found.
We refrain from proceeding further on the analysis of the $U(1)$ charges
since the Konishi anomaly approach is precisely a systematic way to 
obtain the results of that analysis.

There are 6 independent gauge invariants one can build from the matter
fields:
\beq
X_i^{\tilde \imath}=\tilde F^{\tilde \imath}_a 
\epsilon^{}_{bcdef}T^{ab}_{j}T^{cd}_{k}T^{ef}_{i}
\epsilon^{jk}_{},
\label{xii}
\eeq
\beq
Y_i=\epsilon^{}_{\tilde \imath \tilde \jmath}\tilde F^{\tilde \imath}_a 
\tilde F^{\tilde \jmath}_b T^{ab}_{i},
\label{yi}
\eeq
where $\epsilon_{bcdef}$, $\epsilon_{\tilde \imath \tilde \jmath}$ and 
$\epsilon^{jk}$
are the Levi-Civita invariant tensors of $SU(5)$, $SU(2)_{\tilde F}$ and
$SU(2)_T$ respectively.

The above 6 invariants parameterize the 6 dimensional moduli space of the 
theory. In a generic point on the moduli space the gauge group is totally
broken, consistently with the counting that the 24 gauge bosons eat up
24 out of the 30 matter fields, leaving 6 
massless singlets representing the classical flat directions which satisfy
the D-flatness conditions (see \cite{tv} for a complete parameterization
of the solution of these conditions).

Note that $ X_i^{\tilde \imath}$ can be considered as a $2\times 2$ matrix,
and hence $\det X$ is a gauge and flavor symmetry invariant of mass
dimension 8.

According to \cite{ads}, a tree-level superpotential $W_{tree}=\nu Y_1$
lifts all classical flat directions and, together with the non-perturbatively
generated superpotential, leads to dynamical supersymmetry breaking at a scale 
controlled by the coupling $\nu$.

Here we wish to consider a more general tree-level superpotential
given by:
\beq
W_{tree}=\tr \mu X + \nu^i Y_i.
\label{wtree}
\eeq
This is not a renormalizable superpotential, though the first term
can be easily generated  by integrating out some massive
non chiral matter, as is present in 
supersymmetric grand unified models.

Let us first briefly consider
a generalization of the Konishi anomaly which boils down to the classical
equations of motion. For the anomalous one-loop piece of the relation
to be trivially zero, one needs the variation of the field to be independent 
of the field itself. 
It is straightforward to convince oneself that the only non-trivial 
variation satisfying the above requirement is the following:
\beq
\delta \tilde F^{\tilde \imath}_a = \rho^{\tilde \imath ij}
\epsilon_{abcde}T^{bc}_iT^{de}_j, \qquad \delta T^{ab}_i=0.
\label{variation}
\eeq
Acting on the invariants, the variation above gives:
\beq
\delta X^{\tilde \imath}_i=0, \qquad \delta Y_i=-2 \rho^{\tilde \imath jk}
\epsilon_{ij}\epsilon_{\tilde \imath \tilde \jmath}
X^{\tilde \jmath}_k.
\label{genkon1}
\eeq
We have used the fact that there is no gauge invariant built out
of 5 $T$s, and that the only gauge invariant built out of
1 $\tilde F$ and 3 $T$s is in the $\bf 2$ of $SU(2)_T$, and thus is $X$.
Using now the tree-level superpotential, we see that in a supersymmetric
vacuum we must have:
\beq
\nu^i X^{\tilde \jmath}_k=0.
\label{genkon}
\eeq
This means that as soon as the coupling to the $Y_i$ invariants
is turned on at tree-level, the only supersymmetric vacua can occur
at zero value for the $X^{\tilde \imath}_i$ invariants. This is going
to be crucial for the determination of dynamical supersymmetry breaking.

We now turn to determine the ``canonical'' Konishi anomaly relation,
that is the anomaly associated
to the currents generating the variations $\delta \tilde F^{\tilde \imath}_a
=\xi^{\tilde \imath}_{\tilde \jmath} \tilde F^{\tilde \jmath}_a$ and
$\delta T^{ab}_i = \chi_i^j T^{ab}_j$.
Assuming the tree-level superpotential (\ref{wtree}), the following relations
hold in a supersymmetric vacuum:
\beq
(X\mu)^{\tilde \jmath}_{\tilde \imath}
+\delta^{\tilde \jmath}_{\tilde \imath} \nu^i Y_i = 
\delta^{\tilde \jmath}_{\tilde \imath} S,
\label{konf}
\eeq
\beq
(\mu X)^i_j + \delta^i_j \tr \mu X 
+\nu^i Y_j = 3 \delta^i_j S.
\label{kont}
\eeq
The classical equations of motion are recovered by setting $S$ to zero.

By taking traces and substituting back into
the equations, 
it is straightforward to show that the above equations are equivalent to:
\beq
\nu^i Y_j = 0, \qquad (\mu X)^i_j =
\delta^i_j S. 
\label{eom}
\eeq
Considering first the classical vacua (i.e. $S=0$), we notice that 
(\ref{eom}) together with (\ref{genkon}) implies that
as soon as we turn on the $\nu^i Y_i$ coupling, all the flat directions
are lifted since the equations imply that $X^{\tilde \imath}_i=0=Y_i$.
On the other hand, if $\nu^i=0$, not  
all flat directions are lifted, since the expectation values of $Y_i$
remain unconstrained. For instance, take the following values for the matter
fields, which satisfy the D-flatness conditions:
\beq
T^{12}_1=a, \qquad T^{12}_2=b, \qquad \tilde F^{\tilde 1}_1=\tilde 
F^{\tilde 2}_2 = 
\sqrt{|a|^2+|b|^2}.
\label{yconfig}
\eeq
For the above values we have that $Y_1, Y_2 \neq 0$ while 
$X^{\tilde \imath}_i=0$, and one can also check that the classical 
equations of motion are satisfied (for $\nu^i=0$ and $\mu\neq 0$).

\subsection{Unbroken supersymmetry}

We now solve for (\ref{eom}), assuming that $\nu^i=0$:
\beq
X^{\tilde \imath}_i = S
(\mu^{-1})^{\tilde \imath}_i.
\eeq
and use the above result to compute the coupling dependent piece
of the effective superpotential by using the effective relations, 
valid for expectation values in a supersymmetric vacuum:
\beq
X^{\tilde \imath}_i = {\partial W_{eff} \over \partial \mu^i_{\tilde \imath}}.
\eeq
By integrating the above equation, we find:
\beq
W_{eff} = C(S)+S \log (\Lambda^2 \det \mu),
\label{weff}
\eeq
where $C(S)$ is a piece of the superpotential which is independent
of the couplings $\mu^i_{\tilde \imath}$, and $\Lambda$ is the 
holomorphic scale at high energies.

We can now apply the linearity principle \cite{integratein} and integrate
in the effective fields $X$ by subtracting the linear coupling $\tr \mu X$,
to obtain:
\beq
W_{eff}=C(S)-2 S(1-\log {S\over \Lambda^3}) - S \log {\det X \over \Lambda^8},
\eeq
where the procedure is obviously unaffected by the form of $C(S)$.

To find $C(S)$, we use the following argument.
We have to match the above result to the expected low-energy superpotential
of the theory where now $W_{tree}=0$. The invariants are expected
to take on generic expectation values, and break entirely the gauge
group. We thus expect the following $S$-dependent 
Veneziano-Yankielowicz-type  \cite{VY} superpotential:
\beq
W_{eff} = S(1-\log{S\over \Lambda^3} ) - S \log {\det X\over \Lambda^8 }.
\label{vy1}
\eeq

At this point one may question the validity of introducing the gluino
condensate $S$ in (\ref{vy1}) since there is no unbroken low energy
gauge group. However, this is a situation similar to $SU(N_c)$ SQCD
with $N_f=N_c-1$ flavors,\footnote{Indeed, the total matter index, which
plays the role of $N_f$ in our case, is
$n=2\cdot{1\over 2}+2\cdot{3\over 2}=4=5-1$.} 
where a non-perturbative superpotential
is still generated by the instantons in the broken gauge group, and 
correspondingly one can introduce a glueball superfield $S$ and its related
Veneziano-Yankielowicz superpotential as explained for instance
in \cite{aivw} (see also \cite{cdsw}).

By matching with the above, we find:
\beq
C(S)= 3S(1 -\log {S\over \Lambda^3}) .
\eeq
We can thus write the complete effective superpotential (\ref{weff}) as:
\beq
W_{eff} = 3S(1 -\log {S\over \Lambda^3})
+S \log \det (\Lambda^2 \mu) .
\label{wefftot}
\eeq
We will see that the coefficient of 3 is the number of vacua of the theory
with $W_{tree}=\tr \mu X$.

Minimizing (\ref{vy1}) with respect to $S$, we get that:
\beq
S={\Lambda^{11} \over \det X},
\label{sdetx}
\eeq
The non-perturbatively generated superpotential is thus:
\beq
W_{np}={\Lambda^{11} \over \det X}.
\label{wnp}
\eeq
This is nothing else than the superpotential determined by Affleck, Dine
and Seiberg \cite{ads} using symmetry arguments.
Recall that it is generated by a one-instanton contribution (indeed $3N-n=11$)
as it should be in a theory with $N-n=1$. For yet another alternative
way to derive this superpotential, see \cite{rel}.

Now that we have derived $W_{np}$ we can check the vacuum structure of:
\beq
W_{eff}=W_{tree}+W_{np}=\tr \mu X + 
{\Lambda^{11} \over \det X}.
\eeq
Extremizing the above with respect to $X$, we find:
\beq
W_{eff}|_{extr}=3 {\Lambda^{11} \over \det X}, \qquad 
\mbox{with} \qquad \left(\det X \over \Lambda^8\right)^3={1\over \Lambda^2
\det \mu}.
\eeq
Note that the constraint on $\det X$ is cubic, resulting
in 3 vacua. Of course since there is no coupling for the $Y_i$s, all these
vacua are additionally labeled by two flat directions.

Another way to determine that there must be 3 vacua for $\nu^i=0$ is the
following. When $W_{tree}=\tr \mu X$ the generic situation is that
$Y_i$ have non trivial expectation values. Then classically we can take 
for instance a configuration like (\ref{yconfig}). These background matter
fields break the gauge group from $SU(5)$ to $SU(3)$. Out of the 30 matter
superfields, 16 are eaten by the gauge bosons which become massive, while
two remain massless but are neutral, since they parameterize the flat 
directions. The remaining 12 matter fields fit into 2 fundamental and 2
anti-fundamental representations of $SU(3)$. Thus the effective gauge theory
is $SU(3)$ SQCD with $N_f=2$. Moreover $W_{tree}$ gives a tree level
mass to the quarks, so that the low-energy theory is pure $SU(3)$ SYM, 
which has 3 supersymmetric vacua.
This is also an additional justification for the Veneziano-Yankielowicz
piece in (\ref{vy1}).

Let us now briefly comment on a slightly different route, inspired by
\cite{binor}. We could add to the tree level superpotential a piece
like $W_{tree}=\dots +\lambda \det X$, so that the classical equations
of motion would allow for a (frozen) expectation value for $X$, though
only when $\nu^i=0$.
The solution to the Konishi anomaly equations including the above higher
order term would consist of two branches, one which classically reduces
to the origin and the other to the (possibly large) expectation value.
This way one forces the theory to be in a specific vacuum where 
the gauge group is broken and a low-energy superpotential like (\ref{vy1})
is expected. Then by analytic continuation one recovers (\ref{wefftot}),
with additional terms depending on $\lambda$.
The vacuum structure of such a theory consists of 4 vacua, the 3 which
exist for $\lambda=0$ and the additional one with classical
broken gauge symmetry, which is pushed to infinity in the vanishing
$\lambda$ limit.

\subsection{Dynamical supersymmetry breaking}
Let us now finally come to supersymmetry breaking. To recapitulate the 
situation for $\nu^i=0$, we have seen that it is quite similar to
SQCD with $N_f<N_c$: For $W_{tree}=0$, the non-perturbative superpotential
is of runaway behavior (though in this case (\ref{wnp}) is still flat
in the $Y^i$ directions), thus pushing all supersymmetric vacua
to infinity. When $W_{tree}=\tr\mu X$, we recover 3 supersymmetric vacua, 
all being parameterized by 2 additional flat directions.

Now turning on the coupling $W_{tree}=\nu^i Y_i +\dots$ we see that it
immediately leads to supersymmetry breaking: From the classical variation
(\ref{genkon}), we learn that $X$ must vanish. But from the reasoning
of the previous subsection, we had learned that $W_{np}$ is precisely
singular at the origin. Now, by virtue
of the linearity principle $W_{np}$ must be independent of the couplings in
$W_{tree}$, and is thus exact also for $\nu^i\neq 0$.
The outcome is that $W_{eff}$ has no extremum, and thus there is no
supersymmetric vacuum in the theory.
However, since the $\nu^i Y_i$ term lifts all flat directions, while
$W_{np}$ is singular at the origin, the potential must have a minimum
at a finite value of the fields, which will be the non-supersymmetric
vacuum of the theory \cite{ads}.

Note in this respect that if $W_{tree}=\tr\mu X + \nu^i Y_i$, and the $\nu^i$s
are kept small, we expect three minima of the potential close to the
original values of the 3 supersymmetric vacua (without flat directions
of course), with the supersymmetry breaking presumably lifting the degeneracy
among them. On the other hand, if a $\lambda \det X$ coupling is added,
we expect the 4th vacuum to be lifted at a much higher energy than the others
since the $X\neq 0$ vacuum is already ruled out classically.

We have thus established dynamical supersymmetry breaking by deriving 
through the Konishi anomaly first the non perturbative superpotential
and then checking that an additional coupling leads to a total effective
superpotential with no supersymmetric vacua.

An alternative way to show supersymmetry breaking still exploiting
the Konishi anomaly would have been to use (\ref{genkon}) and (\ref{eom})
to show that $Y_i=X^{\tilde \imath}_i=S=0$ in a supersymmetric vacuum.
Then one could have argued as in \cite{ads} that in such a vacuum the
global $U(1)$ symmetries are unbroken, and thus the effective fields
must satisfy the 't~Hooft anomaly matching conditions, which is possible
only at the price of an extremely odd effective field content.
Or in the spirit of \cite{mv} one could have computed a one-instanton
contribution to a correlator involving the above invariants, and upon 
finding a non-zero result the conclusion is that, for instance, $S\neq 0$,
which implies that supersymmetry is broken.

\section{$SU(N)$ chiral gauge theories with one family}
In this section we will consider ${\cal N}=1$ supersymmetric gauge
theories with gauge group $SU(N)$ ($N\geq5$)
with one matter field in the antisymmetric
representation $T^{ab}$ and $N-4$ matter fields in the anti-fundamental
representation $\tilde F^i_a$, $i=1,\dots, N-4$. The matter content is
such that the gauge anomaly cancels.

For all $N\geq 6$, we have the following type of invariants:
\beq
Y^{ij}=\tilde F^i_a\tilde F^j_b T^{ab},
\eeq
which are in the antisymmetric representation of the flavor group
$SU(N-4)$. For $N$ even, we have an additional invariant which
is the Pfaffian of the antisymmetric tensor:
\beq
Z=\epsilon_{a_1\dots a_N}T^{a_1 a_2}\dots T^{a_{N-1}a_N}={\,\rm Pf\,}T.
\eeq
Note that no invariant at all can be written for the $SU(5)$ theory
with one family, which thus does not have any flat direction
nor the possibility to have a tree-level superpotential.

The simplest tree-level superpotential is thus:
\beq
W_{tree}=\nu_{ij}Y^{ij} + \lambda Z.
\label{wtreen}
\eeq
Note that the second term is present only for $N$ even, and is 
non-renormali\-za\-ble for $N>6$.

We can now write the Konishi anomaly relations, for the 
expectation values in a supersymmetric vacuum:
\beq
2\nu_{ik}Y^{kj}=\delta_i^j S ,
\label{mteq}
\eeq
\beq
\nu_{ij}Y^{ij} + {N\over 2} \lambda Z = (N-2) S,
\label{nteq}
\eeq
the second term on the left hand side of (\ref{nteq}) being present
only if $N$ is even. 

\subsection{Odd $N$}

We immediately see that if $N$ is odd, 
the only solution is $Y^{ij}=S=0$ (in the $SU(5)$ case, 
the only solution is $S=0$ since also the first term of $W_{tree}$ cannot
be written).

In this case, we do not dispose of the additional $X$-type invariants as in the
situation with two families, so that we cannot argue as in the previous
section. However the situation is also clearly different: for instance
in the $SU(5)$ case, no invariant can be written and hence no $W_{np}$
either can be written. We have thus to resort to the arguments of
\cite{ads,mv} and say that the conditions $Y^{ij}=S=0$ are not consistent
either with a credible low-energy spectrum, or with one-instanton
calculations.

At this point we comment on the chiral ring of these
theories. It turns out that its structure is very 
simple (somewhat like in SQCD), as it is generated only by the two 
invariants: 
\beq
     S = -{1\over 32 \pi^2}\mathrm{tr}W^\alpha W_\alpha, \quad\hbox{and}\quad 
     Y^{ij} = \tilde{F}^i_a  \tilde{F}^j_b T^{ab},\label{generators}
\eeq
(the second invariant being of course zero for $SU(5)$.)

To see this, consider the chiral ring relations:
\beqs
     W_{\alpha c}^a W_{\beta b}^c &=&- W_{\beta c}^a W_{\alpha b}^c,
     \nonumber \\
     W_{\alpha a}^b \tilde{F}^i_b &=& 0, \label{ring}\\
     W_{\alpha c}^a T^{cb} &=&  W_{\alpha c}^b T^{ca}. \nonumber 
\eeqs

We can construct singlets using the primitive invariants
\beq
     \delta^a_b, \quad \epsilon^{a_1\dots a_N}\quad\hbox{and}
     \quad \epsilon_{a_1\dots a_N}.
\eeq
Let us first look at all the tensor structures that can be constructed
with $\delta^a_b$ only, up to the relations (\ref{ring}). 
They are, in addition to the fundamental fields and the two invariants
(\ref{generators}):
\beqs
     F^{ai} &\equiv& \tilde{F}_b^i T^{ab}, \nonumber \\
     S^a_b &\equiv& \epsilon^{\alpha\beta} W_{\alpha c}^a W_{\beta b}^c,
     \nonumber \\
     T_\alpha^{ab} &\equiv& W_{\alpha c}^a T^{cb} = T_\alpha^{ba},
     \nonumber \\
     \Sigma^{ab} &\equiv& \epsilon^{\alpha\beta} 
     W_{\alpha c}^a W_{\beta d}^c T^{db} = - \Sigma^{ba}. \label{structures}
\eeqs
It is easy to convince oneself employing (\ref{ring}) that no 
more independent structures can
be constructed by simple contraction of the indices and also
that the tensors (\ref{structures}) have the indicated symmetry properties.
Gauge singlets must then be constructed by contracting the above with
the $\epsilon$-tensors. Since 
$\epsilon_{a_1\dots a_N} \epsilon^{b_1 \dots b_N}$ is a product of
$\delta$'s, one can use either all $\epsilon$ in the fundamental or all 
$\epsilon$ in the antifundamental. Using $\epsilon^{a_1 \dots a_N}$
will not work because of the ``lack of indices" (recall that $i=1\dots N-4$) 
so the only possibility is to use $\epsilon_{a_1 \dots a_N}$ which 
restricts the possible tensors that can be used to the two singlets in
(\ref{generators}) and $F^{ai}$, $T^{ab}$ (one of the fundamental
fields), $T^{ab}_\alpha$ and $\Sigma^{ab}$.

Now recall that $T^{ab}$ has one zero eigenvalue, so that by an $SU(N)$
transformation we can take
$T^{aN} = 0$. We then use the chiral ring relations to show that
also 
\beq
     F^{Ni} = T^{aN}_\alpha = \Sigma^{aN} = 0.
\eeq
Thus no further invariants can be constructed with the
$\epsilon$-tensor which always carries a $a=N$ index and we are left
with (\ref{generators}) as the only possibility.
This shows that no non-trivial generalized Konishi anomaly relations 
can be constructed for these theories.

It would be interesting to study the chiral rings for $N$ even and 
for theories with more families. Though the above argument is not
applicable, we believe that their structure is the same, i.e. the only
generators are the glueball superfield and the basic invariants.

\subsection{Even $N$}

On the other hand, when $N$ is even and $\lambda\neq 0$, the Konishi
relations (\ref{mteq}), (\ref{nteq}) can be solved to give:
\beq
Y^{ij}={1\over 2}S (\nu^{-1})^{ij}, \qquad Z={S\over \lambda}.
\label{yzexp}
\eeq
This case can be solved as in the previous section (see also \cite{binor}
for $SU(6)$), finding the
expected non-perturbative superpotential. In this case however, we have
already all the invariants in $W_{tree}$, and we can check that there
is no additional relation like (\ref{genkon}). Hence, there is no
contradiction and all the supersymmetric vacua found in this way
are not lifted, supersymmetry being unbroken.

Let us quickly review how to derive $W_{np}$ and the number of
vacua in this case. From (\ref{yzexp}), we find 
straightforwardly, for $N=2k$:
\beq
W_{eff}=C(S) + S \log (\Lambda^{k-3} \lambda ) +{1\over 2}S \log \det \nu
= C(S) + S \log (\Lambda^{k-3} \lambda {\,\rm Pf\,} \nu  ).
\eeq
We now proceed to integrate in the effective fields $Y^{ij}$ and $Z$,
by subtracting $W_{tree}$ (\ref{wtreen}) 
and solving (\ref{yzexp}) for the invariants.
The result is:
\beq
W_{eff}=C'(S) - (k-1)S(1-\log {S\over \Lambda^3}) -S\log {Z {\,\rm Pf\,} Y
\over \Lambda^{4k-6}},
\label{weffn}
\eeq
where we absorbed in $C'(S)$ a trivial term linear in $S$.
But here we know what the coefficient of the $S\log S$ piece should be:
when $W_{tree}=0$, the invariants take arbitrary values and the gauge
group is maximally broken. By analyzing the D-flatness conditions, 
one finds that the effective theory is always a pure $Sp(4)$ SYM 
with massless neutral fields parameterizing the flat directions.
Hence the overall coefficient of the Veneziano-Yankielowicz term in 
(\ref{weffn})
must be 3. In turn this implies that the exact effective superpotential
is given by:
\beq
W_{eff}=(k+2)S(1-\log{S\over \Lambda^3} ) + S \log (\Lambda^{k-3}
\lambda {\,\rm Pf\,} \nu),
\eeq
and that there should be $k+2$ supersymmetric vacua.

This is confirmed if we integrate out $S$ in (\ref{weffn}), after having
substituted for $C'(S)$, to find that:
\beq
W_{np}=3\left({\Lambda^{2N+3} \over Z {\,\rm Pf\,} Y}\right)^{1\over 3}.
\eeq
Extremizing then $W_{eff}=W_{tree}+W_{np}$ with respect to $Y^{ij}$ and
$Z$ one again finds $k+2$ solutions. 
Of course, 
if we take $\lambda=0$, we have a runaway behavior and no stable vacuum.

\section*{Acknowledgments}

We would like to thank Vanicson L.~Campos for discussions.
This work is partly supported by EU contract HPRN-CT-2000-00122.

\end{document}